\pdfoutput=1
\documentclass[a4paper]{article}

\usepackage{INTERSPEECH2022}
\usepackage{amsmath}
\usepackage{commath}
\usepackage{amssymb}
\usepackage{mathtools}
\usepackage{bbm}
\usepackage{physics}
\usepackage{siunitx}
\usepackage{trfsigns}
\usepackage{pifont}
\usepackage{etoolbox}
\usepackage{multirow}
\usepackage{csquotes}
\usepackage[shortlabels]{enumitem}  
\usepackage[nosort]{cite}   
\usepackage{hyperref}
\usepackage[capitalize]{cleveref}
\usepackage{xcolor}
\usepackage{listofitems}
\usepackage{svg}
\usepackage{balance} 
\usepackage{algpseudocode}
\usepackage{algorithm}
\usepackage{booktabs}

\Crefname{section}{Sec.}{Secs}  %

\usepackage{url}

\newcommand{\cmark}{\ding{51}}
\newcommand{\xmark}{{--}}

\newcolumntype{H}{>{\setbox0=\hbox\bgroup}c<{\egroup}@{}}   

\newcommand*{\thl}{\fontseries{b}\selectfont}
\robustify\thl
\robustify\fontseries

\robustify\underline

\newcommand{\y}{\mathbf{y}}
\newcommand{\z}{\mathbf{z}}
\newcommand{\h}{\mathbf{h}}
\newcommand{\n}{\mathbf{n}}
\newcommand{\B}{\mathbf{B}}

\newcommand{\p}[1]{\mathrm{p}\left(#1\right)}

\DeclareMathOperator*{\argmax}{argmax}

\renewcommand{\H}{^\mathrm{H}}

\newcommand{\inv}{^{-1}}

\usepackage{pgfplots} 
\newlength\fheight 
\newlength\fwidth 
\usepackage{tikz}
\pgfplotsset{compat=1.9}

\usetikzlibrary{external}

\usetikzlibrary{arrows}
\usetikzlibrary{patterns}
\usetikzlibrary{backgrounds}
\usetikzlibrary{fit}
\usetikzlibrary{positioning}
\usetikzlibrary{shapes.geometric}
\usetikzlibrary{calc}
\usetikzlibrary{shapes.multipart}
\usetikzlibrary{shapes.misc}
\usetikzlibrary {shapes.symbols}

\tikzset{>=stealth}

\tikzstyle{block}=[
    draw,
    text depth=0pt,
    thick, 
    rectangle,
    text centered,
    minimum height=3.4ex,
    fill=black!6,
    ] 

\tikzstyle{branch}=[{circle,inner sep=0pt,minimum size=0.3em,fill=black}]
\tikzstyle{box}=[rectangle, rounded corners, draw=black, line width=1pt, text width=2cm]

\tikzstyle{arrow}=[{}-{>}, thick]
\tikzstyle{reverse arrow}=[{<}-{}, thick]

\tikzset{
	do path picture/.style={%
		path picture={%
			\pgfpointdiff{\pgfpointanchor{path picture bounding box}{south west}}%
			{\pgfpointanchor{path picture bounding box}{north east}}%
			\pgfgetlastxy\x\y%
			\tikzset{x=\x/2,y=\y/2}%
			#1
		}
	},
	sin wave/.style={do path picture={    
			\draw [line cap=round] (-3/4,0)
			sin (-3/8,1/2) cos (0,0) sin (3/8,-1/2) cos (3/4,0);
	}},
	cross/.style={do path picture={    
			\draw [line cap=round] (-2/5,-2/5) -- (2/5,2/5) (-2/5,2/5) -- (2/5,-2/5);
	}},
	plus/.style={draw, circle, do path picture={    
			\draw [line cap=round] (-3/5,0) -- (3/5,0) (0,-3/5) -- (0,3/5);
	}},
	mic/.style={inner sep=0pt, do path picture={
			\draw (0,0) circle (0.9);
			\draw [line cap=round] (-0.9, -0.9) -- (-0.9, 0.9);
	}},
	mux/.style={trapezium, draw}
}

\usepackage[acronym,shortcuts]{glossaries}
\glsdisablehyper
\newacronym{SDR}{SDR}{Signal-to-Distortion Ratio}
\newacronym{CSS}{CSS}{Continuous Speech Separation}
\newacronym{PIT}{PIT}{Permutation Invariant Training}
\newacronym{uPIT}{uPIT}{Utterance-level Permutation Invariant Training}
\newacronym{MSE}{MSE}{Mean Squared Error}
\newacronym{DFS}{DFS}{Depth First Search}
\newacronym[]{BLSTM}{BLSTM}{Bidirectional Long-Short-Term Network}
\newacronym{DPRNN}{DPRNN}{Dual-Path Recurrent Neural Network}
\newacronym{WER}{WER}{Word Error Rate}
\newacronym{SSE}{SSE}{Sum Squared Error}
\newacronym{DP}{DP}{Dynamic Programming}
\newacronym{ASDR}{A-SDR}{Averaged \gls{SDR}}
\newacronym{SISDR}{SI-SDR}{Scale-Invariant \gls{SDR}}
\newacronym{SASDR}{SA-SDR}{Source-Aggregated Signal-to-Distortion Ratio}
\newacronym{SASISDR}{SA-SI-SDR}{Source-Aggregated Scale-Invariant \gls{SDR}}
\newacronym{CISDR}{CI-SDR}{Convolution Invariant \gls{SDR}}
\newacronym{SACISDR}{SA-CI-SDR}{Source-Aggregated Convolution Invariant \gls{SDR}}
\newacronym{ASR}{ASR}{Automatic Speech Recognition}
\newacronym{VAD}{VAD}{Voice Activity Detection}
\newacronym{tSDR}{tSDR}{thresholded \gls{SDR}}
\newacronym{logMSE}{log-MSE}{logarithmic Mean Squared Error}
\newacronym{ORCWER}{ORC WER}{Optimal Reference Combination Word Error Rate}
\newacronym{NN}{NN}{Neural Network}
\newacronym{ORCLEV}{ORC Levenshtein Distance}{Optimal Reference Combination Levenshtein Distance}
\newacronym{STFT}{STFT}{Short-Time Fourier Transform}

\newacronym{cACGMM}{cACGMM}{complex Angular Central Gaussian Mixture Model}
\newacronym{cACG}{cACG}{complex Angular Central Gaussian}

\newacronym{EM}{EM}{Expectation Maximization}
\newacronym{SMM}{SMM}{Spatial Mixture Model}
\newacronym{MM}{MM}{Mixture Model}

\newacronym{SAD}{SAD}{Speaker activity detection}
\newacronym{IoU}{IoU}{Intersection-over-Union ratio}
\newacronym{WPE}{WPE}{Weighted Prediction Error}

\newacronym{RIR}{RIR}{Room Impulse Response}
\newacronym{RTF}{RTF}{Relative Transfer Function}

\newacronym{PDF}{PDF}{probability density function}
\newacronym{MVDR}{MVDR}{Minimum Variance Distortionless Response}
\newacronym{wMPDR}{wMPDR}{weighted Minimum Power Distortionless Response}
\newacronym{ICA}{ICA}{Independent Component Analysis}

\newacronym{cpWER}{cpWER}{concatenated minimum-permutation Word Error Rate}

\title{
An Initialization Scheme for Meeting Separation with Spatial Mixture Models
}
\name{Christoph Boeddeker, Tobias Cord-Landwehr, Thilo von Neumann, Reinhold Haeb-Umbach}
\address{Paderborn University, Germany}
\email{\{boeddeker, cord, vonneumann, haeb\}@nt.upb.de}

\begin{document}

\makeatletter
\makeatother

\setlength{\textfloatsep}{8pt plus 0.0pt minus 2.0pt}
\setlength{\floatsep}{8pt plus 0.0pt minus 2.0pt}

\setlength{\abovecaptionskip}{1ex}
\setlength{\belowcaptionskip}{0ex}

\maketitle
\begin{abstract}
Spatial mixture model (SMM) supported acoustic beamforming  has been extensively used for the  separation of simultaneously active speakers. However, it has hardly been considered for the separation of meeting data, that are characterized by long recordings and only partially overlapping speech. In this contribution, we show that the fact that often only a single speaker is active can be utilized for a clever initialization of an SMM that employs time-varying class priors.
In experiments on LibriCSS we show that the proposed initialization scheme achieves a significantly lower Word Error Rate (WER)  on a downstream speech recognition task than a random initialization of the class probabilities by drawing from a Dirichlet distribution. With the only requirement that the number of speakers has to be known, we obtain a WER of \SI{5.9}{\percent}, which is comparable to the best reported WER on this data set.
Furthermore, the estimated speaker activity from the mixture model serves as a 
diarization based on spatial information.

\end{abstract}


\section{Introduction}
Acoustic beamforming guided by \glspl{SMM} has been widely used for blind speech separation given a multi-channel input \cite{Vu2010blind,Ito2016cACGMM,Ito2013permutation,Araki2016MeetingSMM,Araki2018meeting,Boeddeker2018GSS}. 
Building on the sparsity and w-disjoint orthogonality of speech in the \gls{STFT} domain, the vector of channels of a microphone array, which contain the speech of multiple simultaneously active speakers, can be modeled by an \gls{SMM}. Its parameters  are estimated with the \gls{EM} algorithm, of which both batch and frame-online variants exist. The resulting class posterior probabilities are used to estimate spatial covariance matrices, for each speaker in the mixture. From those, the coefficients of statistically optimal acoustic beamformers, such as the \gls{wMPDR} beamformer \cite{Nakatani2019ConvBF}, can be computed, that extract the signals of the individual speakers in the mixture.

Despite being very effective, \glspl{SMM} have hardly been used on meeting data, with \cite{Araki2016MeetingSMM,Araki2018meeting} being notable exceptions.
With meeting data we here denote comparatively long (\SI{1}{minute}) recordings of conversations among a fixed number of participants, which contain segments of no speech activity, of a single speaker talking or multiple speakers being active concurrently.
Meeting separation is concerned with mapping such speech data to output streams in a way such that there is no overlapping speech on those streams. 
The prevalent approach to this so-called \gls{CSS} are neural networks \cite{Chen2020LibriCSS, Chen2021CSSConformer, Chen2021EarlyExit, Wang2021CSSSpectralMapping, Raj2021Meeting}. 
However, an \gls{SMM} has the crucial advantage that its parameters are estimated in an unsupervised fashion, while the supervised neural network training needs both the separate utterances of the individual speakers as targets and their artificial superposition as input to the network. For example, in the CHiME-5 and CHiME-6 challenges \cite{Barker2018CHiME5,Watanabe2020CHiME6}, \glspl{SMM} were the prevailing approach to source separation, because ground truth single-speaker utterances were not available for training.

However, parameter estimation in an \glspl{SMM} is notoriously sensitive to initialization. This is particularly crucial because multiple \glspl{SMM} have to be learnt, namely one for each frequency bin. 
Depending on the initialization, the EM algorithm may settle on an unfavorable local optimum of the likelihood function where the performance of the resulting beamformer is unsatisfactory.

Initialization of \glspl{SMM} can be achieved by either giving initial values to its parameters or to its latent variables, i.e.\ the posterior class affiliation probabilities.
If prior information about the spatial or temporal properties of the recording is available, it can be used to initialize the parameters of the mixture model, as was done, e.g.\ in \cite{Araki2016MeetingSMM} for spatial and \cite{Boeddeker2018GSS} for temporal initialization.
When initializing the class posteriors instead, the most uninformative initialization of the class posterior probabilities is to draw their initial values  at random from a Dirichlet distribution.
But often, some prior knowledge is available to improve over this.
For example, it is often valid to assume that there are noise-only periods at the beginning and end of a recording \cite{Drude2020Thesis}.
The initial values of the posterior of the noise class can then be set accordingly. 

But in case of meeting data, where often only a single speaker is active, one can do better:
We suggest splitting the meeting data into short segments.
Assuming that each segment only contains a single speaker, a single-component \gls{PDF} per frequency is fit to the data of each segment.
The resulting models are clustered to $K+1$ clusters, one cluster for each of the $K$ speakers and one for the noise. 
From each of those clusters a frequency-independent class posterior is computed, which serves as initialization of the \gls{SMM}.

We conducted experiments on the LibriCSS database \cite{Chen2020LibriCSS}, which consists  of  partially overlapping utterances of the LibriSpeech corpus \cite{Panayotov2015Librispeech} that have been played back in a meeting room and recorded with a 7-element microphone array. Using the above sketched initialization scheme we were able to achieve a \gls{WER} of \SI{5.9}{\percent} with a downstream recognizer, which is superior to the \SI{13.3}{\percent} obtained by random initialization and
which comes close to the \SI{5.1}{\percent} \gls{WER} obtained by an oracle initialization. 

\section{Signal model}\label{sec:signalModel}
Let  $\y_{t, f} \in \mathbb{C}^M$ denote the vector of the $M$ microphone signals  in the \gls{STFT} domain, where the indices $t$ and $f$ are the time-frame and frequency-bin index, respectively.
The observation $\mathbf{y}$ contains the sum of $K$ speaker signals
$s_{t, f, k}$, multiplied by their acoustic transfer function $\h_{f, k}$, and noise $\n_{t, f}$.
Those signals are assumed to have the same length as $\mathbf{y}$, i.e.\ the speaker signal contains zeros if a speaker is inactive.
While, in theory, all speakers could be active simultaneously and for the complete duration of the meeting, databases of real conversations \cite{Carletta2005AMI,Barker2018CHiME5} show only partial activity of each speaker.
This partial activity can be modelled by an additional speaker-dependent activity $a_{t, k} \in \{0, 1\}$ without loss of generality, resulting in the model
\begin{align}
    \y_{t, f} = \sum_{k=1}^{K} a_{t, k} \h_{f, k} s_{t, f, k} + \n_{t, f} \label{eg:signalModel}
\end{align}
for the observation signal.


\section{Spatial Mixture Model}\label{sec:smm}

\begin{figure}[t]
    \centering
    \input{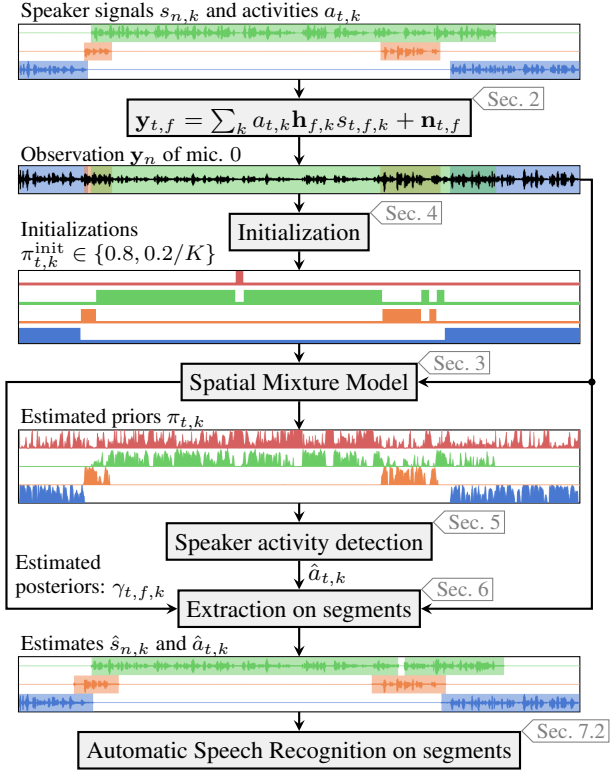}
        
    \caption{System overview, with images from one LibriCSS example. The x-axis represents time (sample $n$ or frame $t$) and the y-axis the values and speaker index. For visualization the permutation of the initialization was removed.}
    \label{fig:overview}
\end{figure}

To model the multichannel observation, we use an \gls{SMM}, in particular the \gls{cACGMM} \cite{Ito2016cACGMM}:
\begin{align}
    \p{\z_{t, f}} &= \sum_{k=1}^{K + 1} \pi_{t, k} \mathcal{A}\left(\z_{t, f}; \B_{f, k}\right) ~~ \text{with} ~~ \z_{t, f} = \frac{\y_{t, f}}{\lVert\y_{t, f}\rVert}
\end{align}
where $\pi_{t,k}$ is a time-varying a priori probability as proposed in \cite{Ito2013permutation}.
Using time dependent priors, instead of frequency dependent ones, better fits a meeting scenario because they can reflect the activity of the speakers over time.
The number of classes $K+1$ is set to the number of speakers $K$ plus one for the noise.
$\mathcal{A}(\cdot)$ is the \gls{cACG} distribution
\begin{align}
    \mathcal{A}(\z_{t, f}; \B_{f, k}) = \frac{(M-1)!}{2\pi^M \det(\B_{f, k})} \left(\z_{t, f}\H \B_{f, k}\inv \z_{t, f}\right)^{-M}
\end{align}
and $\B_{f, k}$ is a hermitian parameter matrix of this distribution.
$(\cdot)\H$ and $(\cdot)\inv$ denote the conjugate transpose operation and matrix inversion, respectively.

The parameters of the \gls{SMM} are trained with the \gls{EM} algorithm, which consists of alternating E- and M-steps. For the \gls{cACGMM} the update equations are \cite{Ito2016cACGMM,Ito2013permutation}: \\
E-step:
\begin{align}
    \gamma_{t, f, k} = \frac{ \pi_{t, k} \det(\B_{f, k})\inv (\z_{t, f}\H \B_{f, k}\inv \z_{t, f})^{-M}}{ \sum_{\kappa = 1}^{K+1}\pi_{t, \kappa}\det(\B_{f, \kappa})\inv (\z_{t, f}\H \B_{f, \kappa}\inv \z_{t, f})^{-M}} \label{eq:Gamma_update}
\end{align}
M-step:
\begin{align}
    \pi_{t, k} &= \frac{1}{F} \sum_{f=1}^{F} \gamma_{t, f, k} \\
    \B_{f, k} &= \frac{M}{\sum_{t=1}^{T} \gamma_{t, f, k}} \sum_{t=1}^{T} \gamma_{t, f, k} \frac{\z_{t, f}\H \z_{t, f}}{\z_{t, f}\H \B_{f, k}\inv \z_{t, f}}, \label{eq:B_update}
\end{align}
where $\gamma_{t, f, k}$ is the a posterior probability that the class $k$ is active at a time-frequency bin $t, f$.
While the time-varying a priori probability \cite{Ito2013permutation} was introduced to prevent the permutation problem between frequencies, it was observed \cite{Drude2020Thesis} that a permutation solver after each \gls{EM} step is nevertheless beneficial.
Here, we use a similarity-based permutation solver.
It is based on the work by Tran Vu \cite{Vu2010blind} and is similar to \cite{Sawada2010PermSolver}.

From now on, we will use the terms prior and posterior as abbreviation for the a priory probability $\pi_{t,k}$ and a posteriori probability $\gamma_{t, f, k}$, respectively.

\section{Initialization}\label{sec:init}
The EM algorithm converges to a local optimum of the objective function and is thus susceptible to its initialization. Either the model parameters ($\pi_{t,k}$, $\B_{f, k}$) or the posteriors ($\gamma_{t, f, k}$) can be initialized. Note, however, that there is no closed-form update equation for the parameter matrix $\B_{f, k}$ of the \gls{cACG}, see \cref{eq:B_update}. Thus, an initialization with the posteriors additionally requires an initial value for  $\B_{f, k}$.
Here, the identity matrix is chosen for the initialization of $\B_{f, k}$.
In the following, several strategies for the initialization of the mixture model are investigated.

\subsection{Oracle initialization}\label{sec:oracle-init}
For the oracle initialization, we follow \cite{PB2018CHiME5} by initializing the posteriors  with the true speaker activities $a_{t,k}$, see \cref{eg:signalModel}, as
\begin{align}
    \gamma_{t, f, k}^{\text{init}} = \frac{a_{t, k}}{\sum_{k=1}^{K+1} a_{t, k}},
\end{align}
The noise class is set to be always active, i.e.\ $a_{t,K+1} = 1$.

Although this initialization uses oracle information,
the \gls{SMM} could forget the initialization while iterating, which makes this scheme different from the guided source separation proposed in \cite{PB2018CHiME5}.

\subsection{Draws from Dirichlet distribution}
The most straightforward initialization of the posteriors is to draw their values from  the Dirichlet distribution as in \cite{Drude2019SMSWSJ}: 
\begin{align}
    \gamma_{t, f, 1}^{\mathrm{init}}, \dots, \gamma_{t, f, K+1}^{\mathrm{init}} \sim \operatorname{Dirichlet}( \boldsymbol{\alpha})
\end{align}
where the $(K+1)$-dimensional parameter vector $\boldsymbol{\alpha}$ of the Dirichlet distribution is set to one for each class.
Instead of sampling the initialization values independently for each frequency, another option is to tie them across the frequency axis, i.e.\ to use the same set of drawn class posteriors for all frequencies.
This frequency-tied initialization can be interpreted as initializing the prior $\pi_{t,k}$ instead of the posterior.
Since $\B_{f, k}$ is initialized as the identity matrix, \cref{eq:Gamma_update} simplifies and the first E-step becomes:
\begin{align}
    \pi_{t, 1}^\text{init}, \dots, \pi_{t, K+1}^\text{init} &\sim \operatorname{Dirichlet}(\boldsymbol{\alpha}) \\
    \gamma_{t, f, k}^\text{init} &= \pi_{t, k}^\text{init} \quad \forall f.
\end{align}

\subsection{Proposed cACG clustering-based initialization}
Using the same value for all elements of the parameter vector $\boldsymbol{\alpha}$ of the Dirichlet distribution leads to an initialization that assumes that all speakers are active for roughly the same amount of time.

However, one can do better.
In a typical meeting, most speakers are at least once the sole active speaker.
For each of these single-speaker regions, a single \gls{cACG} is sufficient to model  the observations. 
We use this property to find the following clustering-based initialization. 
First, the meeting is split into short segments. Then, the parameters of a  \gls{cACG} are estimated for each of these segments.
If two segments, $\ell$ and  $l$, have the same active speaker, the estimated parameter matrices $\B_{f, \ell},  \B_{f, l}$ of these segments will exhibit a low distance to each other.
To measure their distance, the correlation matrix distance \cite{Herdin2005correlation} is used:
\begin{align}
    d(\B_{f, \ell}, \B_{f, l}) = 1 - \frac{\tr\left( \B_{f, \ell}, \B_{f, l} \right)}{\left\lVert\B_{f, \ell}\right\rVert_\mathrm{Fro}  \left\lVert\B_{f, l}\right\rVert_\mathrm{Fro}}
\end{align}
where $\left\lVert\cdot\right\rVert_\mathrm{Fro}$ and $\tr(\cdot)$ are the Frobenius norm and trace operator, respectively.
Using this distance metric, the pair-wise distance matrix $\mathbf{D}$ with the entries $D_{\ell, l}$
\begin{align}
    D_{\ell, l} = \frac{1}{F} \sum_{f=1}^{F} d(\B_{f, \ell}, \B_{f, l})
\end{align}
is constructed.
Based on this distance matrix, all similar segments are grouped together to obtain an initialization for the $K+1$ classes. 
This grouping is achieved by complete linkage clustering \cite{Sorensen1948CompleteLinkageClustering}, a variant of agglomerative hierarchical clustering (AHC).
The clustering starts with each segment forming its own  cluster. Next, two clusters are merged if the maximum distance between entries of the one and the other cluster is the smallest compared to all other potential cluster merges. This process is repeated until $K+1$ clusters remain. 
We used the complete linkage clustering because it keeps the cluster diameter small, so that it is unlikely that it groups two speakers in one cluster.
We then initialize
\begin{align}
    \pi_{t, k}^{\text{init}} &= \begin{cases}
        0.8 & \text{if segment } \ell \text{ of } t \text{ belongs to } k  \\
        0.2/K & \text{otherwise.} \\
    \end{cases}
\end{align}
\cref{fig:overview} illustrates an example for this initialization, where
30 consecutive frames are used to estimate $\B_{f, \ell}$.






\section{Speaker activity detection}\label{sec:sad}
After the convergence of the \gls{EM} algorithm, speaker activity is detected through the learned class prior probabilities $\pi_{t, k}$.
This process is divided into three steps: noise class identification, class fusion, and speech segmentation. Note that this delivers the diarization information of which speaker is active and when. 

All three steps are done with thresholded dilation and/or erosion operations.
Both are well known techniques from image processing that apply as follows to the 1-dimensional case of speech signals:
For a dilation, a sliding window is moved over the signal with a shift of one, and the maximum value inside the window is the new value for its center sample.
In erosion, we do the same, just with the minimum operation.

\subsection{Noise identification}
We observed that a class prior $\pi_{t, k}$ is close to zero when the corresponding speaker is inactive, and the noise is active all the time (See estimates in \cref{fig:overview}).
We can estimate the active time of a class by smoothing the class prior $\pi_{t, k}$ with a dilation followed by an erosion and then discretizing with a threshold.
Thus, the noise class $k^\mathrm{noise}$ is identified as the class that is the most active:
\begin{align}
    \bar{\pi}_{t, k} = \operatorname{Erode}(\operatorname{Dilate}(\pi_{t, k}, 101), 101) \\
    k^\mathrm{noise} = \argmax_k \left\{ \smash{\sum_t}\vphantom {\sum} \left(\bar{\pi}_{t, k}  > 0.2 \right) \right\}
\end{align}
where $101$ is the frame length of the sliding window.

\subsection{Class fusing}
Sometimes, an initialization leads the \gls{SMM} to use two classes for one speaker while another speaker is ignored.
To mitigate this error, the mixture classes that represent the same speaker are fused. This fusion is again based on the prior $\pi_{t, k}$ which is smoothed with dilation and erosion. If the \gls{IoU}
\begin{align}
    \mathrm{IoU}(k, \kappa ) = \frac{\sum_t \left(\bar{\pi}_{t, k}  > 0.2 \right) \wedge \left(\bar{\pi}_{t, \kappa}  > 0.2 \right)}{\sum_t \left(\bar{\pi}_{t, k}  > 0.2 \right) \vee \left(\bar{\pi}_{t, \kappa}  > 0.2 \right)} \label{eg:IoU}
\end{align}
between two classes $k$ and $\kappa$ is larger than $0.8$, they are fused, i.e.\ their priors and posteriors are summed up.

\subsection{Speech segmentation}
Finally, we estimate the speech activities $\hat{a}_{t, k}$ from $\pi_{t, k}$, i.e.\ smooth and discretize $\pi_{t, k}$. Here, only the dilation is applied to overestimate the activity instead of accidentally removing speech regions at the beginning and end of each utterance.
Initial and final silence is no issue for an \gls{ASR} system and this further ensures that the segments are long enough for the \gls{ASR} system to operate on.
To keep the false alarms rate small, we had to select a larger value for the threshold as was chosen for noise detection and speaker fusion:
\begin{align}
    \hat{a}_{t, k} = \operatorname{Dilate}(\pi_{t, k}, 79) \ge 0.5.
\end{align}
This estimate $\hat{a}_{t, k}$ can be interpreted as a diarization based on spatial information.


\section{Extraction}\label{sec:extraction}
The speech extraction is performed on segments, defined by continuous activity in $\hat{a}_{t, k}$, where $k$ is the target speaker.
%
To extract the target speech from one segment, a mask-based convolutional beamformer \cite{Nakatani2019ConvBF}, factorized \cite{Boeddeker2020FactorizedConvBF,Nakatani2020ConvBF} into a \gls{WPE} dereverberation \cite{Nakatani2010WPE} and a \gls{wMPDR} beamformer \cite{Nakatani2019ConvBF} component, is used.
%
As the target mask for a segment of speaker $k$, the posterior $\gamma_{t, f, k}$ of this segment is used.
For the distortion mask, the remaining speakers, floored by a small value $\max({\sum_{\kappa \ne k}}\gamma_{t, f, \kappa}, 10^{-4} )$ are used.
Because of space limitations, we skip the equations here. They can be found in \cite{Boeddeker2020FactorizedConvBF} and \cite{Nakatani2020ConvBF}.

\begin{table*}[t!]
  \caption{cpWER on LibriCSS segments for different initializations, fusions, and w/ and w/o \gls{WPE} dereverberation as preprocessing.
  }
  \label{tab:resultsSMM}
  \centering
  

    \sisetup{round-precision=1,round-mode=places, table-format = 2.1}

  \begin{tabular}{c l c c c 
    H
    | S S S S S S |
    H
    S }
    \toprule
    & {\textbf{Initialization}} & \multicolumn{2}{c}{\textbf{Class fusion}} & {\textbf{WPE}} & {EM} &  \multicolumn{6}{c|}{\textbf{Overlap ratio in \%}} & {Missed} & {\textbf{Average}} \\
    & & {intermediate} & {final} & & {iterations} & {$0$S} & {$0$L} & {$10$} & {$20$} & {$30$} & {$40$} & {speaker} & {} \\
    \midrule
    (1) & Dirichlet & \xmark & \xmark & \xmark & 100 & 11.671435383881736 & 11.126587464154035 & 13.538168793373629 & 15.3014486578611 & 22.89239204934887 & 21.099756234764673 & 0 & 16.562211895737686 \\  
    (2) & Dirichlet (frequency-tied) & \xmark & \xmark & \xmark & 100 & 13.60276585598474 & 12.216304793117574 & 14.39444817550929 & 17.52236898167874 & 23.200822481151473 & 22.487342958934935 & 0 & 17.852918990901724 \\  
    (3) & Dirichlet & \xmark & \cmark & \cmark & 100 & 11.337625178826896 & 11.118394100778369 & 10.113051264830982 & 14.26288879420537 & 15.651620483697249 & 15.91505719107444 & 337 & 13.3503440954341 \\  
    \midrule
    (4) & Proposed (frequency-tied) & \xmark & \xmark & \xmark & 100 & 5.5317119694802095 & 6.366243342892257 & 6.794269084396687 & 9.554750745632724 & 11.367864486438853 & 15.71348209263079 & 0 & 9.672853244927037 \\  
    (5) & Proposed (frequency-tied) & \xmark & \xmark & \cmark & 100 & 5.215784453981879 & 4.702990577632118 & 5.641370047011417 & 8.201959948870899 & 10.599236267502203 & 15.422838927432963 & 0 & 8.802138141046534 \\ 
    (6) & Proposed (frequency-tied) & \xmark & \cmark & \cmark & 100 & 4.840247973295184 & 4.662023760753789 & 4.661965524960824 & 7.008947592671495 & 8.508763340840106 & 10.861616351021938 & 112/3421 & 7.06350167158675 \\ 
    (7) & Proposed (frequency-tied) & \cmark & \cmark & \cmark & 100 & \thl 4.16070577014783 & \thl 4.301515772224499 & \thl 4.225430937989702 & \thl 5.725394120153387 & \thl 7.279937334769411 & \thl 8.50834427151697 & 65 & \thl 5.918068968086196 \\ 
    \midrule
    (8) & Oracle (frequency-tied) & \xmark & \xmark & \xmark & 20 & \color{gray} 3.9163090128755367 & \color{gray} 3.990167963949201 & \color{gray} 4.208641146183121 & \color{gray} 5.3259480187473365 & \color{gray} 6.501517673553314 & \color{gray} 6.117569848115507 & 0 & \color{gray} 5.1488596891500515 \\  
    \bottomrule
  \end{tabular}

\vspace{-2ex}
\end{table*}

\begin{table}[th]
\centering
  \caption{Literature comparison.
  \cite{Chen2020LibriCSS,Chen2021CSSConformer,Wang2021CSSSpectralMapping} reported the asclite WER (no diarization) and we calculated the average WER from the individual ones. \cite{Raj2021Meeting} reported cpWER with source counting.
  }
    \label{tab:resultsOther}
    
    
    \sisetup{round-precision=1,round-mode=places, table-format = 2.1}
    \setlength\tabcolsep{2.8pt} 
  \begin{tabular}{ l | S S S S S S | S }
    \toprule
    \textbf{System} & \multicolumn{6}{c|}{\textbf{Overlap ratio in \%}} & {\textbf{Average}} \\
    & {$0$S} & {$0$L} & {$10$} & {$20$} & {$30$} & {$40$} & \\
    \midrule
    Chen et al. \cite{Chen2020LibriCSS} & 11.9 & 9.7 & 13.6 & 15.0 & 19.9 & 21.9 & 15.983124889414524 \\
Raj et al. \cite{Raj2021Meeting} & 11.2 & 8.5 & 10.6 & 14.8 & 15.5 & 17.5 & 13.4 \\
    Chen et al. \cite{Chen2021CSSConformer} & 5.2 & \thl 4.0 & 5.8 & 6.8 & 9.0 & 10.0 & 7.119566411815651 \\
    Wang et al. \cite{Wang2021CSSSpectralMapping} & 5.4 & 5.0 & 4.8 & \thl 5.0 & \thl 6.6 & \thl 7.6 & \thl 5.850056340389075 \\
    Proposed & \thl 4.16070577014783 & 4.301515772224499 & \thl 4.225430937989702 & 5.725394120153387 & 7.279937334769411 & 8.50834427151697 & \thl  5.918068968086196 \\  
    \bottomrule
  \end{tabular}
  
\end{table}

\section{Experiments}

\subsection{Dataset}
For the experiments, we use the LibriCSS database \cite{Chen2020LibriCSS} with a sample rate of \SI{16}{\kilo\hertz}.
This database is meant to simulate a realistic meeting scenario:
Eight loudspeakers were placed in a conference room and played back utterances from the LibriSpeech corpus \cite{Panayotov2015Librispeech} with a variable amount of overlap between the speakers.
A seven-channel  microphone array recorded the playback.
To investigate different conversation styles, the database has different setups: short silence (0S), long silence (0L) and different overlaps between \SI{10}{\percent} and \SI{40}{\percent}.
For each setup, 10 meetings with a duration of \SI{10}{minutes} were created.
Afterwards, the data was split at silence to obtain smaller segments with an average length of \SI{50}{seconds} containing 1 to 8 speakers.
%


\subsection{Automatic Speech Recognition}\label{sec:asr}
\glsreset{cpWER}
\glsreset{ASR}

We perform \gls{ASR} with a pretrained LibriSpeech end-to-end speech recognizer \cite{watanabe2020PretrainedASR} from the ESPnet framework \cite{Watanabe2018_ESPnetEndtoEndSpeech}.
It uses a transformer architecture with an RNN-based language model and achieves a \gls{WER} of \SI{2.6}{\percent} on the clean test set of LibriSpeech.
We did not adapt the model to separation artifacts.

To assess the performance, we use the \gls{cpWER} \cite{Watanabe2020CHiME6}, which is computed as follows.
All transcriptions from the same speaker are concatenated for both the reference and the estimation.
Since the ordering of the references and estimates are usually arbitrary, the lowest \gls{WER} among the permutations between references and estimates is reported.

This is different from other works on LibriCSS \cite{Chen2020LibriCSS,Chen2021CSSConformer,Wang2021CSSSpectralMapping}.
Often, the separation systems produce two output streams without diarization. 
In this case, it is not possible to obtain speaker-wise transcriptions.
So, alternatively, the transcriptions from the same output stream are concatenated and the asclite tool \cite{github-usnistgov-SCTK}, which uses temporal information to align multiple reference utterance transcriptions with the estimated stream transcriptions, is used to compute a (speaker agnostic) \gls{WER}.




\subsection{Results}
\Cref{tab:resultsSMM} shows our results on LibriCSS, where 100 \gls{EM} iterations are used and the \gls{STFT} size and shift are 1024 and 256, respectively.
In row (1) the same initialization as in \cite{Drude2019SMSWSJ} is used.
Changing the initialization to frequency-tied (2), the average performance gets worse.

Replacing the initialization with our proposed initialization (4),
we see a clear improvement and using \gls{WPE} \cite{Drude2018nara} as preprocessing (5) further improves the \gls{cpWER}.
With added class fusion (6) after the EM, the performance is improved, although it discards some speakers.
Finally, adding an intermediate speaker fusion (7), where we start with $K+3$ classes (two more than in (6)) and force a fusion after 10 and 20 \gls{EM} iterations, i.e.\ fuse the classes with the highest IoU (see \cref{eg:IoU}), yields an \gls{cpWER} of \SI{5.9}{\percent}.
As an upper bound on the performance, we initialized the \gls{SMM} with the annotated utterance boundaries (8), see \cref{sec:oracle-init}, and obtained a \gls{cpWER} of \SI{5.1}{\percent}.


Since this is a publically available dataset, we list a few other publications in \cref{tab:resultsOther}.
We could not find other separation systems without \glspl{NN}, hence we can only compare with \glspl{NN}.
The best system that we found that estimates per-speaker transcriptions is \cite{Raj2021Meeting}.
In contrast to us, they also did source counting.
Most other publications \cite{Chen2021CSSConformer,Wang2021CSSSpectralMapping} build on top of the \gls{CSS} idea from \cite{Chen2020LibriCSS}, and hence only separate and recognize speech, but are unable to re-identify speakers.
We see a clear improvement over \cite{Raj2021Meeting}, the LibriCSS baseline \cite{Chen2020LibriCSS} and followup system \cite{Chen2021CSSConformer}.
Our system is on par with
the currently best reported system in \cite{Wang2021CSSSpectralMapping}, albeit with an unsupervised separation approach.



\section{Conclusion} 
We proposed an initialization scheme for a spatial mixture model to cope with meeting separation, where a class density of the \gls{SMM} is learned on small segments and then clustered.
We showed, on LibriCSS, that this initialization is superior to a conventional initialization.
Using dereverberation and an also proposed class fusion, the spatial mixture model is comparable to the best reported WER on this data set, produced by an \gls{NN}-based meeting separation.

\section{Acknowledgements}
Computational resources were provided by the Paderborn Center for Parallel Computing.
Funded by the Deutsche Forschungsgemeinschaft (DFG, German Research Foundation) - Project 448568305.

\balance
\bibliographystyle{IEEEtran}

\bibliography{mybib}

\end{document}